\documentclass[aps,prd,reprint, nofootinbib, superscriptaddress]{revtex4-1}

\usepackage{amsmath,amssymb, amsthm,amstext}
\usepackage{natbib}
\usepackage{graphicx}
\usepackage{color}
\usepackage{array, enumerate}
\usepackage{bm}
\usepackage{multirow}
\usepackage[breaklinks,colorlinks,citecolor=blue]{hyperref}
\usepackage{braket}
\usepackage{txfonts}

\usepackage{soul}
\setulcolor{magenta}
\setul{-2.5pt}{1.5pt}

\def\nn{\nonumber}
\def\be{\begin{equation}}
\def\ee{\end{equation}}
\def\ba{\begin{aligned}}
\def\ea{\end{aligned}}
\newcommand{\I}{\text{I}}

\begin{document}

\title{Superradiant clouds may be relevant for close compact object binaries }
\author{Ao Guo}
\email{guoao23@mails.ucas.ac.cn}
\affiliation{International Centre for Theoretical Physics Asia-Pacific, University of Chinese Academy of Sciences, 100190 Beijing, China}
\author{Jun Zhang}
\email{zhangjun@ucas.ac.cn}
\affiliation{International Centre for Theoretical Physics Asia-Pacific, University of Chinese Academy of Sciences, 100190 Beijing, China}
\affiliation{Taiji Laboratory for Gravitational Wave Universe (Beijing/Hangzhou), University of Chinese Academy of Sciences, 100049 Beijing, China}
\author{Huan Yang}
\email{hyangdoa@tsinghua.edu.cn}
\affiliation{Department of Astronomy, Tsinghua University, Beijing 100084, China}
\affiliation{Perimeter Institute for Theoretical Physics, Ontario, N2L 2Y5, Canada}
\affiliation{University of Guelph, Guelph, Ontario N1G 2W1, Canada}

\begin{abstract}
Bosonic fields (within suitable mass range) may be collectively generated by rotating black holes through the black hole superradiance process. The resulting black hole is surrounded by a "cloud" of particles whose wave function populates the superradiant energy level of the black hole. For comparable mass ratio binary black hole systems, it has been suggested that these clouds often deplete at large binary separations because of level mixing effects. As a result, these clouds may not be dynamically relevant for black hole and neutron star binaries that enter the LIGO-Virgo-KAGRA and LISA detection frequency band. In this work, we point out that the common envelope process during a compact binary evolution may bring the binary to $\sim 0.01$AU in hundreds to thousands of years, so that depletion caused by certain level mixings are no longer important. We derive a relevant regime of binary parameters where the clouds are still present for binary entering the LISA band, and show that common envelop process does enlarge such parameter regime. When the binary separation further decreases due to gravitational wave radiation, we discuss the impact of possible cloud mass transfer between the binary objects. 
\end{abstract}
\maketitle

\section{Introduction}

Ultralight bosons such as axionlike particles (ALPs) are possible candidates for dark matter \cite{Turner:1983he,Press:1989id,Hu:2000ke,Amendola:2005ad,Schive:2014dra,Hui:2016ltb}.
Their mass and coupling strength to the Standard Model particles are constrained by various cosmological, astrophysical, and terrestrial observations and experiments \cite{Graham:2015ouw,Irastorza:2018dyq,marsh2016axion}. In particular, because of the well-known black hole superradiance mechanism, rotating astrophysical black holes ranging from $\sim 10 {\rm M}_\odot\textbf{--}10^9 {\rm M}_\odot$ could superradiantly excite such bosons to form a macroscopic superradiant cloud if the de Broglie wavelength of the bosons is comparable to the size of the black hole. In this case, electromagnetic and gravitational wave observations of properties, environments and distributions of astrophysical black holes can be used to probe/constrain ultralight bosons. For example, recent gravitational wave observations with ground-based detectors have made use of isolated black holes and binary merger remnants are starting to constrain ALPs in the mass range close to $\sim 10^{-12}$eV \cite{Brito:2017zvb,Sun:2019mqb,Isi:2018pzk,Tsukada:2018mbp,Tsukada:2020lgt,Zhang:2021mks,palomba2019direct,yuan2022constraints,Ng:2020ruv,zhu2020characterizing,jones2023methods}.

On the other hand, it is interesting to consider black holes or neutron stars in the inspiral stage, as the presence of superradiant clouds around binary components may change the binary orbital dynamics, in the frequency band relevant for space-borne and/or ground-based gravitational wave detectors. There are indeed both analytical and numerical investigations of binary dynamics in related regimes (e.g., \cite{Ikeda:2020xvt,Baumann:2022pkl,Choudhary:2020pxy,Zhang:2018kib, Zhang:2019eid, Baumann:2019ztm,
Tomaselli:2023ysb}), assuming superradiant clouds are present. However, it was suggested in Refs.~\cite{Baumann:2018vus,Berti:2019wnn} that, if a black hole binary inspirals from large separation because of gravitational wave radiation, the binary generically crosses so-called resonant transitions so that the cloud switches from a superradiant state to a decay state and dissipates in a much shorter time. Although some resonant depletions are shown to be negligible due to nonlinear effects~\cite{takahashi2022axion, Takahashi:2023flk}, the resulting amount of the cloud in the relevant detection band for space-borne and ground-based detectors is often too small to be detectable, at least for comparable mass ratio binaries.  Therefore, the clouds may be absent for premerger stellar-mass binary black holes relevant for gravitational wave observation.

In this work, we point out that, by considering the astrophysical evolution history of stellar-mass binary black holes, there is a significant chance that superradiant clouds are still retained at least in the LISA band. This is because the relevant binaries often involve the common envelope (CE) phase, during which the binary separations greatly decrease to allow relatively short binary inspirals by gravitational wave radiation. Depending on the mass of the boson particles, it is possible that the binary rapidly crosses a resonance during the CE evolution, suppressing the corresponding depletion, and/or resides at a separation that is smaller than some of the depletion bands. In the latter case, the cloud of the less massive black hole may grow and enter the LISA band without experiencing depletion induced by level mixing. We shall consider two events (GW$150914$ and GW$151226$) to illustrate these scenarios, finding a range of particle mass where the superradiant clouds are retained at smaller separations relevant for gravitational wave observations. It is reasonable to expect binary black holes with different component masses may be used to probe/constrain ultralight bosons with different masses. This also motivates further studies on the dynamics of binary black holes with the presence of superradiant clouds.

One interesting feature we find, as the binary black holes inspiral towards each other, is that the superradiant clouds may undergo resonant state transfer between two black holes, so that the cloud organically populated near one black hole may efficiently transfer to the other binary companion at a certain orbital frequencies. This type of cloud transfer is different from Roche-Lobe mass transfer commonly studied in binary stars, as the cloud transfer requires the orbital frequency to match level frequency difference between the two binary components. We work out the condition for such resonant cloud transfer and observe that in a certain parameter regime it also leads to noticeable backreaction on the orbit. Notice that even if one of the binary components is a neutron star that cannot superradaintly generate the cloud efficently by itself, the resonant cloud transfer may be able to reshuffle some of the cloud to the neutron star, offering a viable mechanism of neutron stars carrying clouds in a binary. This is interesting if the particles constituting the clouds weakly couple to the Standard Model particles, in which case additional electromagnetic signatures may be expected. We shall leave discussion in this direction to future studies.

We work with the $(-,+,+,+)$ signature and take $\hbar=c=1$. For the notations, $\mu$ and $\omega$ denote the mass and the frequency of the scalar field, respectively. $M$ and $a$ denote the mass and the dimensionless spin of the superradiance black hole, while $M_*$, $\theta_*$ and $\phi_*$ are the mass, elevation angle, and azimuth angle of the companion star/black hole. When discussing binary evolution, we also use $M_1$ and $M_2$ to denote the mass of the primary and secondary black hole/star in the binary. $R$ and $\Omega$ are the orbital separation and the orbital frequency of the binary system. $f$ is the frequency of GWs emitted by the binary. We further define the gravitational radius $r_g = G M$, the total mass $M_{\rm tot} = M_1+M_2$, and the dimensionless reduced mass $\eta = M_1 M_2/M_{\rm tot}^2$ for the binary, as well as the fine structure constants $\alpha_1= G\mu M_1$ and $\alpha_2= G\mu M_2$ for each black hole. We use $|n\ell m \rangle$ to denote the bound state of superradiant cloud associated with a certain black hole.

\section{Superradiant clouds in progenitor evolution of black hole binaries}

\begin{figure}
    \centering
    \includegraphics[width=0.95\linewidth]{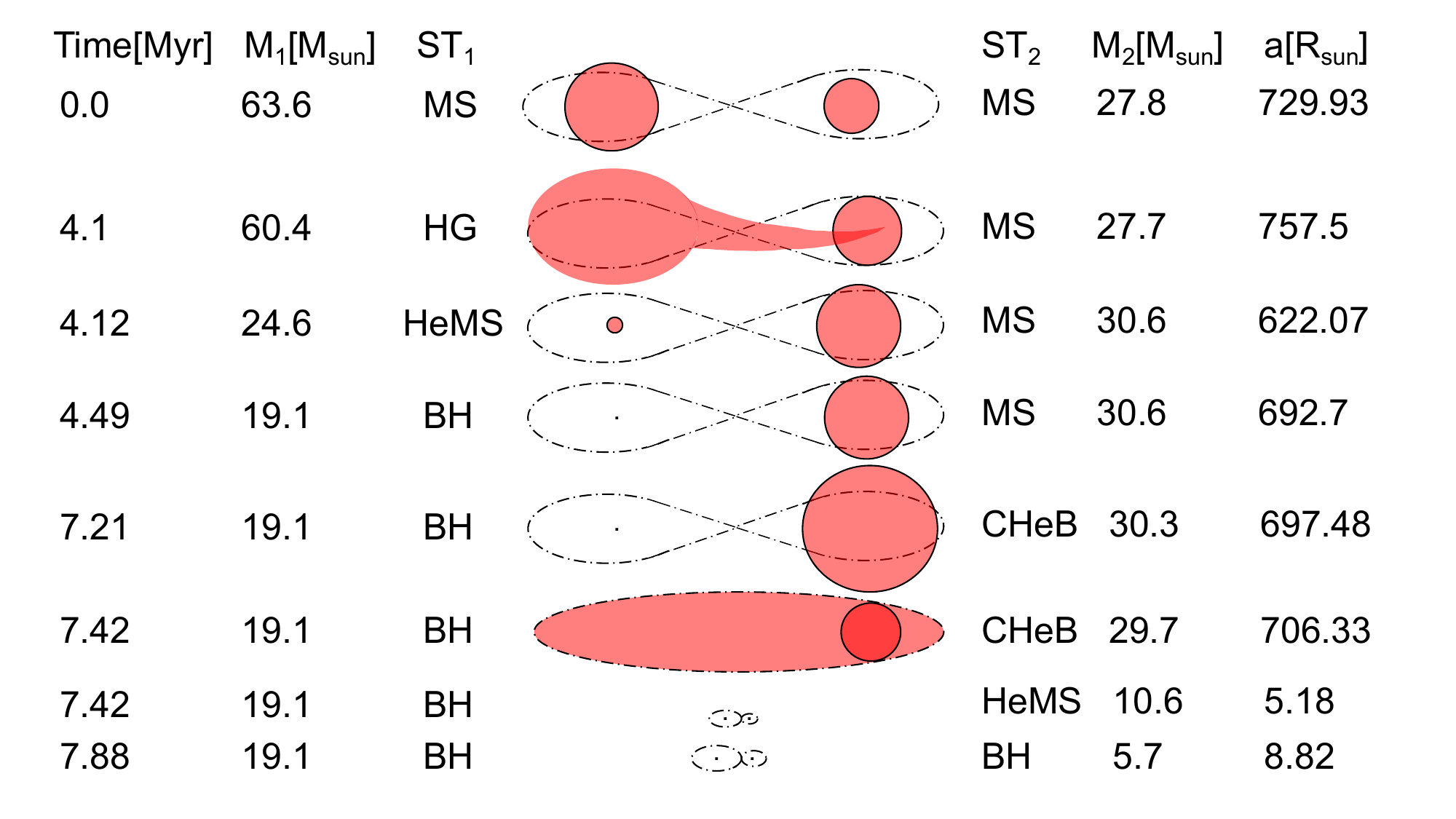}
     \caption{Classical isolated binary evolution of a GW151226-like binary.}\label{fig:his}
\end{figure}

Black hole binaries could form through several evolutionary channels. For demonstration, we shall first consider the progenitor evolution of a GW151226-like binary \cite{LIGOScientific:2016sjg}, and discuss the possible evolution of superradiant clouds in such a binary. The evolutionary pathway of GW151226 has been investigated in Ref.~\cite{Stevenson:2017tfq}, according to which the binary is formed through the classical isolated binary evolution channel. See Fig.~\ref{fig:his} for a cartoon demonstration of the formation history. Specifically, the binary initially has two high-mass main-sequence O stars, a primary of $\sim 64 M_\odot$ and a secondary of $\sim 28M_\odot$ with an initial orbital separation of $\sim 730 R_\odot$. At the end of its main sequence evolution, the primary expands, filling its Roche lobe and initiating mass transfer. Because of the mass transfer and stellar winds, the primary loses $\sim 40 M_\odot$ and eventually forms a black hole of $\sim 19 M_\odot$. The secondary, on the other hand, grows to a star of $\sim 30 M_\odot$, and is $\sim 690 R_\odot$ away from the primary at the formation of the primary black hole. As evolution continues, the secondary also expands and forms a CE enclosing the binary. The CE evolution would significantly harden the orbital, reducing the orbital separation from $\sim 706 R_\odot$ down to $\sim 5 R_\odot$ in a very short time. In the end, the secondary also forms a black hole of $\sim 6 M_\odot$ after $\sim 3.39\ {\rm Myrs}$ evolution from the formation of the primary black hole. Then the formed binary, consisting of a $\sim 19 M_\odot$ black hole and a $\sim 5.7 M_\odot$ black hole at a separation of $\sim 8.82 R_\odot$, would merge and be detected by LVK in $\sim 300\  {\rm Myrs}$.

\begin{figure}
    \centering
       \includegraphics[width=1.0\linewidth]{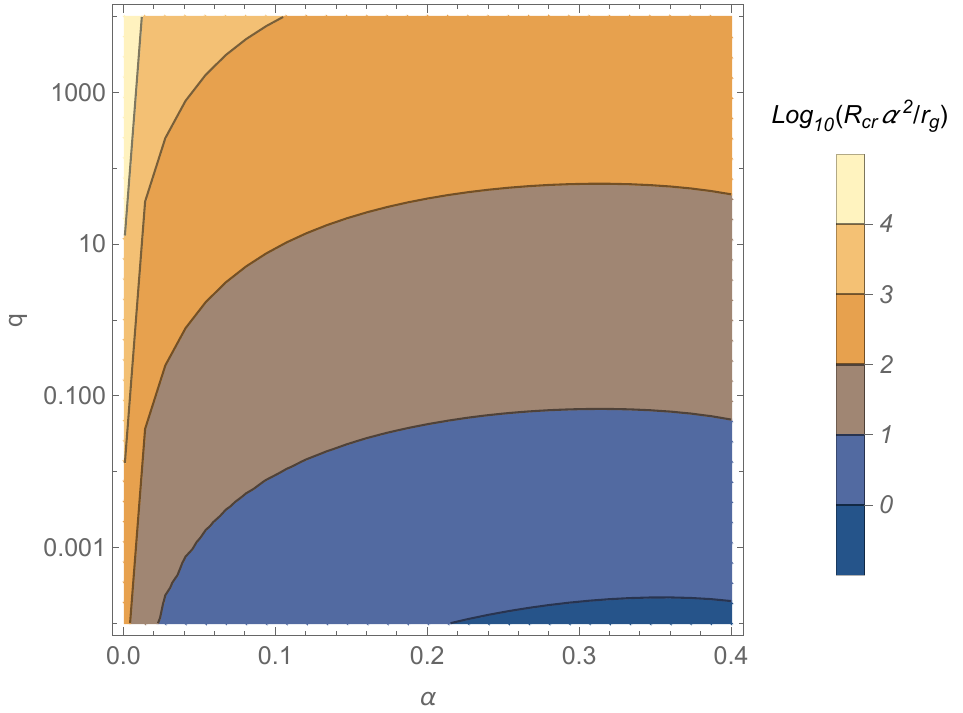}
     \caption{Orbital separation where superradiance is suppressed by tidal interactions. Here the mass ratio is defined as $q\equiv M_*/M$.}\label{fig:src}
\end{figure}

Now we discuss the possible evolution of superradiant clouds in the above binary evolution. There are several timescales that are relevant for the discussion. One is the timescale of the superradiance growth, which can be estimated by the inverse of the growth rate $\Gamma_{n\ell m}$. For free scalar fields around an isolated rotating black hole, the growth rate of the $|n\ell m\rangle$ mode can be estimated by \cite{Detweiler:1980uk, Pani:2012bp, Rosa:2012uz}
\be
\Gamma_{n \ell m} = \frac{2 r_+}{G M} C_{ n \ell m }(\alpha)  \left( m \Omega_{\rm H} - \omega \right) \alpha^{4\ell + 5}
\label{growrate}
\ee
with
\be
\ba
C_{n \ell m}\left( \alpha \right) \equiv & \frac{2^{4\ell +1} (n + \ell)!}{n^{2\ell + 4} (n - \ell- 1)!} \left[ \frac{\ell !}{(2\ell)! (2\ell + 1)!}\right]^2 \nn \\ 
&\prod_{j=1}^{\ell} \left[ j^2 \left( 1 - a^2 \right)  + \left(a m - 2 \tilde{r}_+ \alpha  \right)^2 \right], 
\ea
\ee
for $\alpha \ll 1$.\footnote{Superradiance could be suppressed by self-interactions and coupling to the other particles~\cite{Baryakhtar:2020gao,Spieksma:2023vwl}, which we shall not consider in this paper.} For rotating black holes in binaries, the growth rate can be suppressed due to tidal interactions~\cite{Tong:2022bbl}. In Fig.~\ref{fig:src}, we show the critical separation $R_{\rm cr}$ where superradiance of the $|211\rangle$ mode ceases in the presence of a given companion. In particular, we improve the formula in Ref.~\cite{Tong:2022bbl}, and extend the results to a regime where separation is smaller than the cloud radius (see Appendix~\ref{App:levelmixing} for details).  From Fig.~\ref{fig:src}, we find that for comparable mass ratio binaries ($q =M_*/M\sim 1$), superradiance is suppressed at $\sim 10 r_g/\alpha^2$ for $\alpha \sim 0.1$ and at $\sim 100 r_g/\alpha^2$ for $\alpha \sim 0.01$.

Once formed, a superradiant cloud will gradually lose energy by radiating GWs and eventually deplete. Such GW radiation defines the lifetime of an isolated cloud $\tau_{\rm GW}$. For $|211\rangle$ mode, the lifetime can be estimated by~\cite{Yoshino:2013ofa, Brito:2017zvb,arvanitaki2010string,arvanitaki2015discovering,siemonsen2023modeling}
\be
\tau_{\rm GW} \simeq 10^7 \text{ years} \left( \frac{M}{3M_\odot} \right) \left( \frac{0.07}{\alpha}\right)^{15}\,.
\ee
for $\alpha \lesssim 0.1$. In most of the paper, we shall consider the
case with $\alpha \lesssim 0.1$ so that the above equation and the nonrelativistic approximation are valid. Besides GW radiation, clouds in a binary system could also deplete due to level mixing~\cite{Baumann:2018vus, Berti:2019wnn}, resonantly and/or nonresonantly. For corotating orbits, i.e., the spins of the host black hole and the orbit align, there is hyperfine resonance, which takes place at~\cite{Baumann:2018vus} 
\be
\ba
R_{\rm H} & = 144^{1/3} \, \alpha^{-4} (1+q)^{1/3} \hskip 1pt a^{-2/3} \, r_g \, , \\ 
\ea
\ee
assuming circular orbits. For comparable mass ratio binaries, while the hyperfine resonance may not take place due to nonlinearity~\cite{takahashi2022axion}, a cloud can still deplete nonresonantly as the orbit approaching to $R_{\rm H}$~\cite{Berti:2019wnn}.
In addition to the level mixing discussed in Refs.~\cite{Baumann:2018vus, Berti:2019wnn}, the tidal potential also couples the superradiant mode to decay modes with the $\ell = 0$ and $\ell = 1$ multipoles when the companion is within the exponential tail of the cloud profile (see Appendix~\ref{App:levelmixing} for details). These multipoles can lead to fine and Bohr resonances that take place at $R_{\rm F} \sim \alpha^{-10/3} r_g$ and $R_{\rm B} \sim \alpha^{-2}r_g$ respectively, resulting in efficient depletion of the cloud. Bohr resonance takes place at a separation comparable to the cloud radius, in which case the companion is deeply immersed in the superradiant cloud and the perturbation analysis may not be justified at least for comparable mass ratio binaries. Nevertheless, Bohr level mixing and the corresponding nonresonant depletion start when the companion dip in the exponential tail of the cloud, i.e., at a separation much larger than the cloud radius.

\begin{figure}
    \centering
    \includegraphics[height=0.7\linewidth]{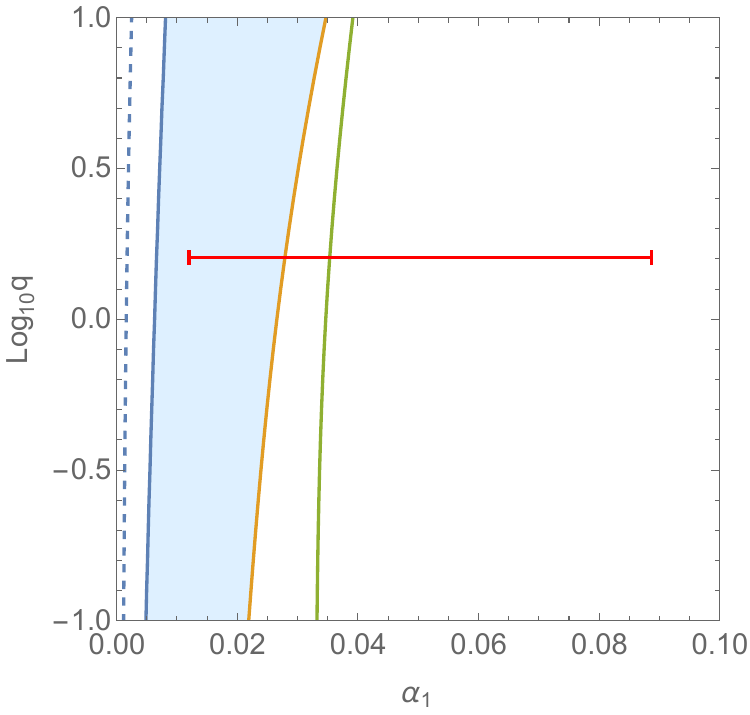}
     \includegraphics[height=0.7\linewidth]{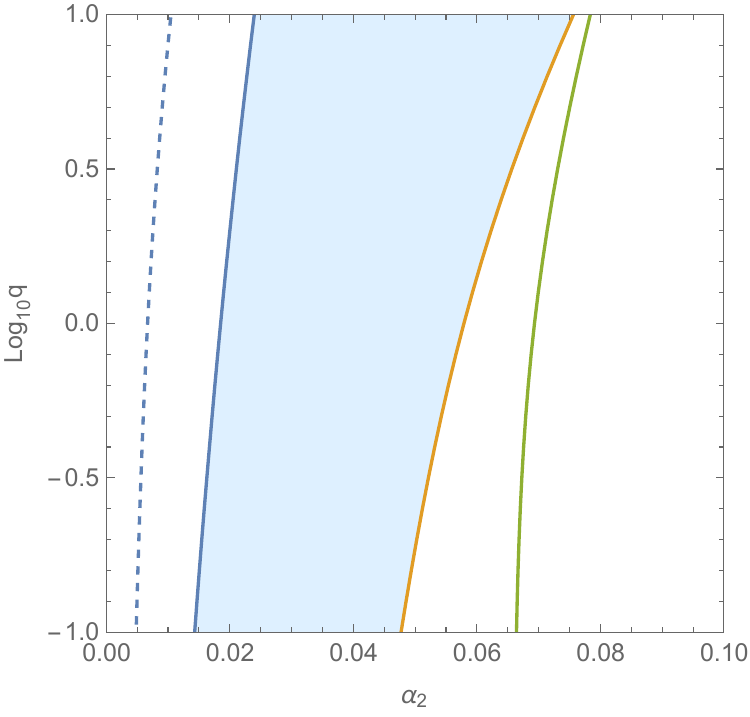}
     \caption{Superradiance parameter space of the $|211\rangle$ mode in a binary. We consider superradiance around a primary black hole of $19.1 M_\odot$ which is initially $692.7 R_\odot$ away from its companion in the upper panel, and that around a secondary black hole of $5.7 M_\odot$ which is initially $8.82 R_\odot$ away from its companion in the lower panel. Here $q = M_*/M$ is defined as the mass ratio between the companion and the superradiance black hole. Namely, $q=M_2/M_1$ in the upper panel and $q=M_1/M_2$ in the lower panel with $M_{1}$ and $M_{2}$ being the mass of the primary and secondary black hole/star, respectively. The dashed blue, solid blue, orange, and green lines saturate the conditions $\tau_{\rm orb}^{\rm GW} \ge 1/\Gamma_{211}$, $R_i \ge R_{\rm cr}$, $\tau_{\rm orb}^{\rm GW} \le \tau_{\rm GW}$, and $R_i \le R_{\rm H}$, respectively. Assuming a circular orbit decaying purely by GW radiation, the blue regions show the parameter space in which the cloud can efficiently grow and survive up to merger before it depletes purely by its own GW radiation. In particle, the orbit could decay faster due to CE evolution, altering the correspond range in $\alpha_1$. For example, the red segment denotes such range, considering the progenitor evolution of the GW151226-like binary. 
     }\label{fig:con}
\end{figure}

Having these scales in mind, we now discuss the possible evolution of the cloud in the progenitor evolution of a GW151226-like binary. We mainly consider the orbital evolutionary pathway of the GW151226-like binary, and assume the black holes formed with relatively high spins to allow cloud excitation. We first focus on superradiance near the primary black hole, which is $\sim 19 M_\odot$ at formation and has a $\sim 30 M_\odot$ companion with an initial separation $R_{i} \sim 690 R_\odot$. A sufficient growth of the cloud around the primary black hole requires at least $\tau_{\rm orb} > 1/\Gamma$, where $\tau_{\rm orb}$ is the lifetime of orbital after the primary black hole formed. Assuming the orbit decays by GW radiation, the orbital lifetime can be estimated by
\be\label{orbtGW}
\tau_{\rm orb}^{\rm GW} = \frac{5}{256} \frac{R_i^4}{q(1+q)G^3 M_1^3},
\ee
where $R_i$ is the initial separation. In a realistic case, the orbital lifetime can be altered by mechanisms such as CE evolution. Take the GW151226-like binary, for example, the CE evolution can reduce the separation from $706 R_\odot$ to $5 R_\odot$ in less than $0.01 \ {\rm Myrs}$, which could take $ 10^{10}\ {\rm Myrs}$ by purely GW radiation. In addition, we find that the superradiance is not suppressed by the tidal interactions for $\alpha > 0.008$ given the relative large separation at the black hole formation, cf.~Fig.~\ref{fig:src}. In Fig.~\ref{fig:con}, we show the parameter space for superradiance growth in a binary. In the upper panel of Fig.~\ref{fig:con}, we consider a primary black hole of $19.1 M_\odot$ formed in a binary with a separation of $692.7 R_\odot$. Assuming a general corotating orbit that decays purely by GW radiation, the $|211\rangle$ mode could sufficiently grow on the right-hand side of the blue line, and shall not deplete up to merger due to its own GW radiation on the left-hand side of the orange line. While in the presence of CE evolution, the corresponding range in $\alpha_1$, for example denoted by the red line for the GW151226-like binary, can be altered. We shall focus on corotating systems in this work, as it is natural to expect that field binaries are more likely to be corotating.

After efficient growth, the cloud in a corotating system can avoid hyperfine depletion if the initial separation is deeply within the hyperfine resonance orbit, i.e., $R_i \ll R_{\rm H}$. While for $R_i > R_{\rm H}$, the cloud may still survive from the hyperfine depletion, if the orbit sweeps through the hyperfine depletion band during the later CE evolution. Because in this case, the orbit decays in a very short time, and the nonresonant hyperfine level mixing may not cause efficient depletion. As orbital separation continues decreasing, the cloud can be further depleted due to the fine level mixing induced by the $\ell=1$ multipole, and then the Bohr level mixing and ionization~\cite{Baumann:2021fkf, Baumann:2022pkl}. A cloud can also deplete due to Bohr resonance if the system is contourrotating~\cite{Baumann:2018vus, Berti:2019wnn}.

\begin{figure}
    \centering
    \includegraphics[height=0.9\linewidth]{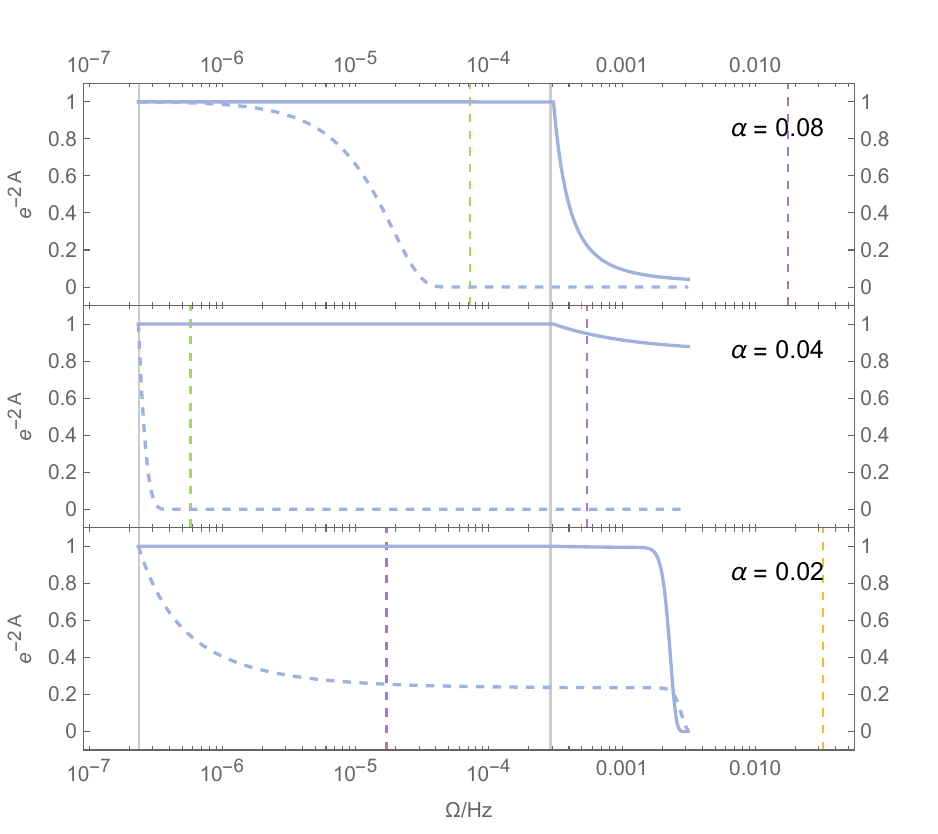}
       \caption{Depletion of the cloud around the primary black hole in the GW151226-like binary. The solid gray lines denote the beginning and the ending of the CE evolution, while the dashed green, purple and yellow lines show the resonance frequencies of the $|21-1\rangle$, $|200\rangle$, and $|100\rangle$ modes, respectively. For comparison, we also show cloud depletion in a similar binary in the absence of CE evolution in blue dashed lines.}
    \label{fig:dep}
\end{figure}

To be concrete, we shall investigate cloud depletion in the GW151226-like binary. In our calculation, we shall consider contributions from $|21-1\rangle$, $|200\rangle$, and $|100\rangle$ modes, as contributions from the other modes are suppressed given the selection rules.\footnote{In principle, the cloud can also deplete due to the ionization process discussed in Ref.~\cite{Baumann:2018vus}. We expect that ionization only becomes efficient at small orbital separations, say $R \lesssim 10 r_g /\alpha^{2}$, and does not affect the scenario considered here.} 
We start from $t_i$, i.e., the time of the primary black hole formation, and divide the orbital evolution into three stages: The pre-CE stage lasts for $2.93$ Myrs, in which the mass ratio is $q=1.60$ and the orbit decays by purely GW radiation, i.e., 
\be
\dot{\Omega} = \frac{96}{5} \frac{\eta}{G^2 M_{\rm tot}^2} (GM_{\rm tot} \Omega)^{11/3}.
\label{eq:dOGR}
\ee
The CE stage lasts for 1000 years, in which the mass ratio is $q=1.55$ and we assume $\dot{\Omega} = \Omega/1000\ {\rm yrs}$.
In the post-CE stage, the mass ratio becomes $q=0.55$, and the orbit decays following Eq.~\eqref{eq:dOGR}. We find that when outside the CE stage, most of the clouds deplete nonresonantly, in which case the mass of the superradiant cloud is approximately proportional to $e^{-2\mathcal{A}(t)}$ with
\begin{equation}
    \mathcal{A}(t)\equiv\sum_{n, \ell }\sum_{m\leq0}|\Gamma_{n \ell m}|\int^{t}_{t_i}dt'|c_{n\ell m}(t')|^2.
    \label{eq:dep1}
\end{equation}
Here $|c_{n\ell m}(t')|^2$ is the amplitude of a certain decaying mode $|n\ell m\rangle$, the evolution of which can be found in Appendix~\ref{App:levelmixing} as well as in Refs.~\cite{Baumann:2018vus,Berti:2019wnn}. 
In Fig.~\ref{fig:dep}, we show the cloud depletion for different $\alpha$. For $\alpha=0.08$, we find the orbit sweeps the $|21-1\rangle$ resonance frequency during its CE stage, and the cloud slightly depletes due to the mixing to the $|21-1\rangle$ mode. After the CE stage, the cloud depletes completely before the orbit enters the milli-Hz band due to the mixing to the  $|200\rangle$ mode. For $\alpha=0.04$, the cloud barely depletes as the orbit sweeps the $|21-1\rangle$ frequency band in its CE stage. It only depletes slightly in the post-CE stage due to the mixing to the $|200\rangle$ mode, and has about $80\%$ of its mass left when the orbit enters the milli-Hz band. For $\alpha = 0.02$, the cloud survives from the $|200\rangle$ depletion due to CE evolution, but depletes completely due to the mixing to the $|100\rangle$ mode. For comparison, we also show cloud depletion without CE process (with dashed lines) in Fig.~\ref{fig:dep}. We can find that the CE process does save clouds from depletion in certain parameter regimes as the orbit passes the depletion band in a very short time during the CE stage.

\begin{figure}
    \centering
    \includegraphics[height=0.7\linewidth]{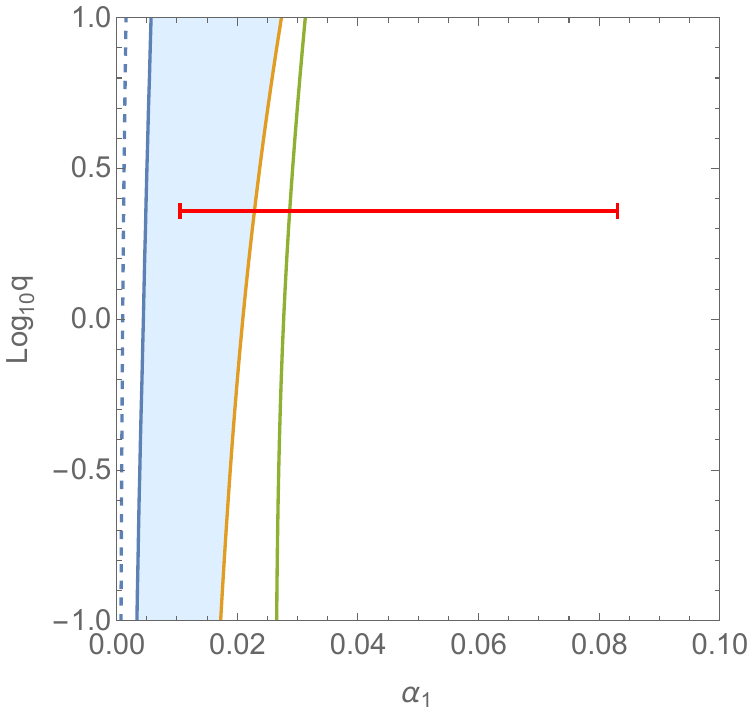}
     \includegraphics[height=0.7\linewidth]{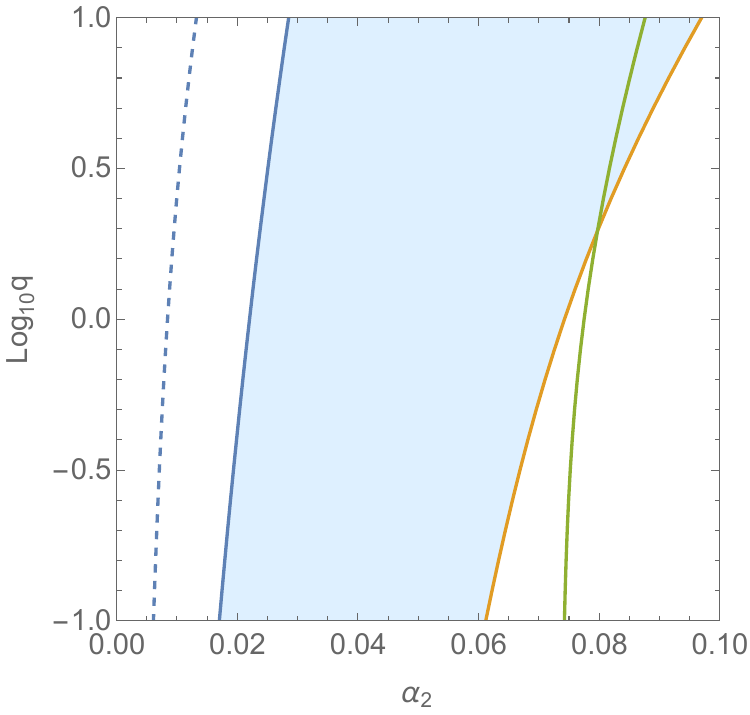}
     \caption{Superradiance parameter space of the $|211\rangle$ mode in a binary. We consider superradiance around a primary black hole of $35 M_\odot$ which is initially $3620.4 R_\odot$ away from its companion in the upper panel, and that around a secondary black hole of $30 M_\odot$ which is initially $25.8 R_\odot$ away from its companion in the lower panel. The labels are the same as those in Fig.~\ref{fig:con}. The red segment denotes range in which the cloud can efficiently grow and survive up to merger before it depletes purely by its own GW radiation, assuming the progenitor evolution of the GW150914-like binary. 
     }\label{fig:con0914}
\end{figure}

A superradiant cloud can also grow around the secondary black hole that forms in the later binary evolution. The same discussion applies, except that there is no CE evolution in the following orbital evolution. The lower panel in Fig.~\ref{fig:con} shows the parameter space for the cloud hosted by the secondary black hole. One may be careful about whether superradiance is terminated by the tidal interactions given the small initial separation. As another example, we also consider the progenitor evolution of a GW150914-like binary. In this case, the primary black hole forms at $35 M_\odot$ and with a companion of $80 M_\odot$ at $3620.4 R_\odot$. $0.35\  {\rm Myrs}$ after the primary black hole formation, the binary undergoes a CE evolution, reducing the orbital separation to $25.8 R_\odot$. After another $0.4 \ {\rm  Myrs}$ evolution, the secondary black hole of $30 M_\odot$ forms and the orbital separation is $ 25.8 R_\odot$. The parameter space of superradiance in such a binary is shown in Fig.~\ref{fig:con0914}.

\section{Implications for binary black holes in milli-Hertz GW observations}

Stellar mass black hole binaries are one of the target sources for space-borne GW detectors, such as LISA, Tianqin and Taiji, which are sensitive to GWs of milli-Hertz. The dynamic signatures of black hole binaries with superradiant clouds have been extensively discussed~\cite{Cardoso:2011xi, Ferreira:2017pth, Baumann:2018vus, Berti:2019wnn, Zhang:2018kib, Zhang:2019eid, Tomaselli:2023ysb,Cao:2023fyv}. Therefore, it is interesting to know whether superradiant clouds, if formed in black hole binaries, could survive to the milli-Hertz band. 

\begin{figure}
    \centering
    \includegraphics[height=0.8\linewidth]{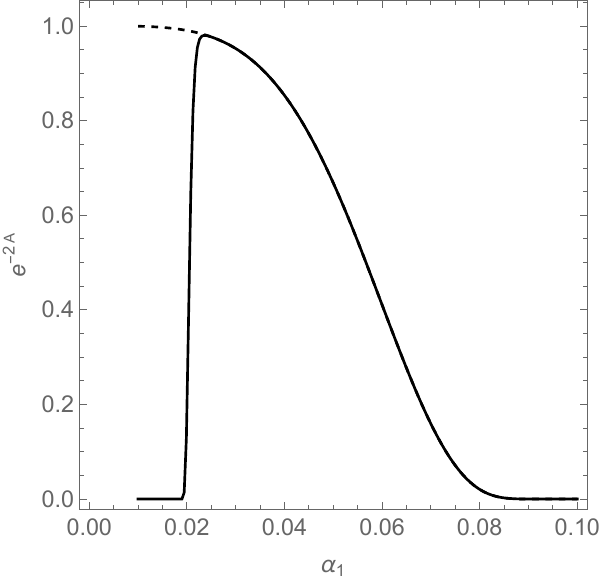}\\
     \includegraphics[height=0.8\linewidth]{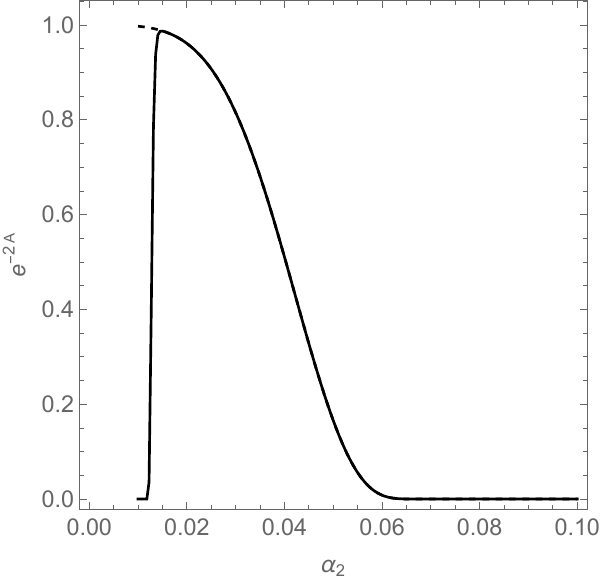}
    \caption{Parameter space for cloud around the primary black hole (upper panel) and secondary black hole (lower panel) in a GW151226-like binary to survive to LISA band. The cloud depletion is calculated with Eq.~\eqref{eq:dep1}, with $t_i$ being the formation time of the host black hole and $t$ being the time when the orbital frequency is about $2 \pi \time 10^{-3} \ {\rm Hz}$. We consider level mixing to the $|21-1\rangle$, $|200\rangle$, and $|100\rangle$ modes, and find that for $\alpha_1 < 0.025$ ($\alpha_2 <0.015$) the cloud depletes completely due to mixing with the $|200\rangle$ mode, while for $\alpha_1 > 0.05$ ($\alpha_2 <0.04$) the cloud depletes significantly due to mixing with the $|21-1\rangle$ mode. In order to show the effects from the $\ell=0$ modes, the dashed lines show the results considering only level mixing with the $|21-1\rangle$ mode.}
    \label{fig:dplisa}
\end{figure}

\begin{figure}
    \centering
    \includegraphics[height=0.8\linewidth]{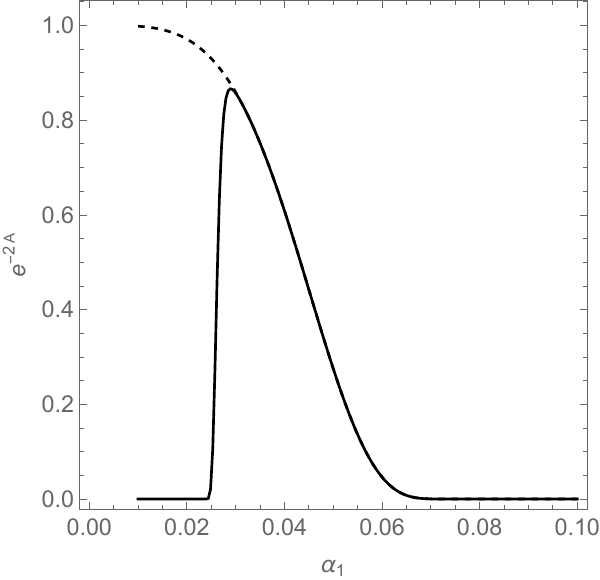}\\
     \includegraphics[height=0.8\linewidth]{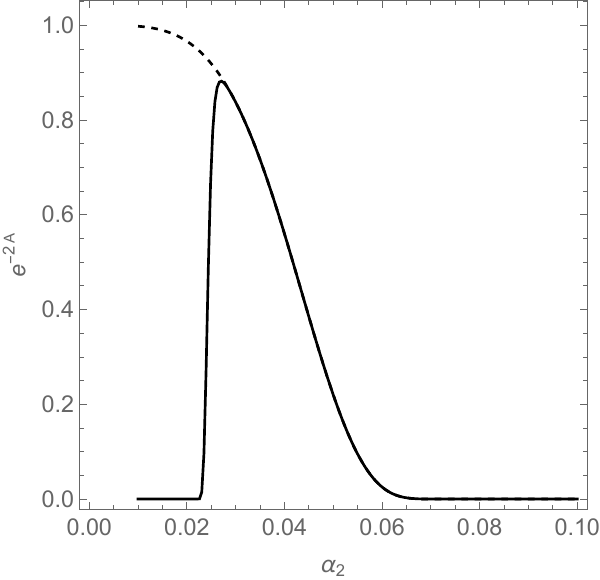}
    \caption{Parameter space for cloud around the primary black hole (upper panel) and secondary black hole (lower panel) in a  GW150914-like binary to survive to LISA band. The notation is similar to that in Fig.~\ref{fig:dplisa}.}
    \label{fig:dplisa0914}
\end{figure}

In Fig.~\ref{fig:dplisa}, we investigate cloud depletion due to level mixing caused by tidal interaction before the binary enters the milli-Hertz band. We consider the progenitor evolution of the GW151226-like binary, and perform the integral~\eqref{eq:dep1} from the formation of the host black hole to the time when the binary enters the milli-Hertz band, i.e., $\Omega/\pi = 10^{-3}\  {\rm Hz}$. For a cloud around the primary black hole, we find that the cloud can survive to the milli-Hz band when $\alpha_1 \sim 0.04$. The cloud will deplete due to nonresonant level mixing with the $|21-1\rangle$ if $\alpha_1>0.07$, and will deplete due to the  mixing with $|200\rangle$ mode if $\alpha_1< 0.025$. For a cloud around the secondary black hole in the GW151226-like binary, the cloud may survive to the LISA band if $0.015 < \alpha_2 <0.04$. The cloud depletion in the GW150914-like binary is shown in Fig.~\ref{fig:dplisa0914}.

For stellar mass black hole binaries, the orbital separation is about $10^3 \textbf{--} 10^5 r_g$ when they are visible in the LISA observation band. For $\alpha \sim 0.1$, the orbital separation is typically much larger than the size of the cloud, and the orbital dynamics can be affected by the finite size effects of the clouds. In particular, the presence of a superradiant cloud inevitably induces extra multipole moments in addition to that of its host black hole. For the $|211\rangle$ mode, the quadrupole moment of the cloud is
\be
Q= -6\alpha^{-4} M_c G^2 M^2,
\ee
where $M_c$ is the mass of the superradiant cloud. Such a quadrupole moment affects the inspiralling waveform at the second post-Newtonian (PN) order~\cite{Ryan:1995wh, Ryan:1997hg,Poisson:1997ha, Pappas:2015npa, Loutrel:2022ant}. In order to estimate the effects on the GW waveform, we consider the number of cycles that GWs spend in a logarithmic frequency interval,
\be
\frac{{\rm d} N }{{\rm d} \ln f} =\frac{f^2}{\dot{f}},
\ee
where we can write $\dot{f} = \dot{f}_{\rm BH} + \dot{f}_{\rm quad}$ with~\cite{Poisson:1997ha}
\be
\dot{f}_{\rm quad} = - \frac{96 \eta Q }{\pi G^2  M_{\rm tot}^2 M^3} \left(G M_{\rm tot} \pi f \right)^{5} 
\ee
Then the additional cycles caused by the quadrupole moment in a logarithmic frequency interval can be estimated as
\be
\Delta N (f) \simeq  \frac{25 Q}{96\pi \eta M G^2 M_{\rm tot}^2} \left( G M_{\rm tot} \pi f \right)^{-1/3}
\ee
On the other hand, the time spent in the a logarithmic frequency interval is
\be
T(f) = \frac{5 G M_{\rm tot}}{256 \eta} \left(\pi G M_{\rm tot} f \right)^{-8/3}.
\ee
Assuming one year observation of LISA, the number of additional cycles in GW phase caused by the cloud's quadrupole moment is
\be
\ba
\Delta N_{\rm obs} &\simeq   \Delta N (f) \frac{T_{\rm obs}}{T(f)}  \nn \\
&\simeq \frac{50}{(1+q)^{2/3}} \left(\frac{M_c}{\alpha M}\right) \left(\frac{M}{30 M_\odot}\right)^{7/3} \left(\frac{f}{10^{-3}\  {\rm Hz}}\right)^{7/3} \left(\frac{\alpha}{0.05}\right)^{-3}
\ea
\ee
which is detectable even if $M_c$ is relatively small. For $\alpha \sim 0.01$, the orbital separation is typically comparable to the size of the cloud, in which case the interplay between binary and the cloud can be complex, and could be detected through its imprints on the GWs emitted by binaries.

\section{Binary Mass Transfer}

A superradiant mode in a binary could not only couple to the other modes of its host black hole due to the tidal interaction, but also the modes of the second black hole and the modes of the binary. While coupling to the modes of its host black hole, bounded and unbounded, leading to level mixing and ionization, we expect that coupling to the modes of the second black hole and the binary could lead to cloud mass transfer and formation of CE of the cloud. In this section, we shall focus on the coupling to the modes of the second black hole and discuss the possible cloud mass transfer.\footnote{There is a related discussion presented in Ref.~\cite{Guo:2023lbv}, which considered equally populated clouds to start with, while here we consider the more realistic scenario that a cloud is initially excited around one of the black holes and discuss the following evolution.}

\subsection{Formalism}

In the nonrelativistic limit, the cloud in a binary satisfies
\begin{equation}
    i \frac{\partial}{\partial t} \psi({\bf r},t)=\left(-\frac{\nabla^{2}}{2\mu}-\frac{\alpha_1}{| {\bf r}- {\bf r}_1|}-\frac{\alpha_2}{|{\bf r}-{\bf r}_2|}\right) \psi({\bf r},t),
\label{eq}
\end{equation}
where ${\bf r}_1$ and ${\bf r}_2$ are the position of the primary and secondary black hole, respectively. We make the following ansatz:
\begin{equation}
   \psi({\bf r},t)= c_{1i}(t)\, \psi_{1i}({\bf r},t)+ c_{2j}(t)\, \psi_{2j}({\bf r},t),
   \label{an}
\end{equation}
where we assumed summation on repeated subscriptions, and $\psi_{ai}$ with $a=1,2$ satisfying
\begin{equation}
    i \frac{\partial}{\partial t} \psi_a({\bf r},t)=\left(-\frac{\nabla^{2}}{2\mu}-\frac{\alpha_a}{| {\bf r}- {\bf r}_a|}\right) \psi_a({\bf r},t),
\end{equation}
is the modes of each black hole, where $i$ labels various modes. Substituting ansatz~\eqref{an} into Eq.~\eqref{eq} and considering the normalization of $\psi_{ai}$, we can solve the equations of $c_{ai}$:
\begin{equation}
\begin{aligned}
i\dot{c}_{1k}& (\delta_{kj}-\langle\psi^*_{2i}\psi_{1k}\rangle\langle\psi^*_{1j}\psi_{2i}\rangle) \\
= -c_{1k}& (\langle\psi^*_{1j}V_{2}\psi_{1k}\rangle-\langle\psi^*_{1j}\psi_{2i}\rangle \langle\psi^*_{2i}V_{2}\psi_{1k}\rangle) \\
-c_{2k}&(\langle\psi^*_{1j}V_1\psi_{2k}\rangle-\langle\psi^*_{1j}\psi_{2i}\rangle\langle\psi^*_{2i}V_1 \psi_{2k}\rangle), 
\label{dotc1}
\end{aligned}
\end{equation}
\begin{equation}
\begin{aligned}
i\dot{c}_{2k}&(\delta_{kj}-\langle\psi^*_{1i}\psi_{2k}\rangle\langle\psi^*_{2j}\psi_{1i}\rangle) \\
=-c_{1k}&(\langle\psi^*_{2j}V_{2}\psi_{1k}\rangle-\langle\psi^*_{2j}\psi_{1i}\rangle \langle\psi^*_{1i}V_{2}\psi_{1k}\rangle) \\
-c_{2k}&(\langle\psi^*_{2j}V_1\psi_{2k}\rangle-\langle\psi^*_{2j}\psi_{1i}\rangle\langle\psi^*_{1i}V_1 \psi_{2k}\rangle),
\label{dotc2}
\end{aligned}
\end{equation}
where $V_a = -\alpha_a/|{\bf r} - {\bf r}_a| +  \alpha_a  {\bf r}\cdot {\bf r}_a/R^3$ is the gravitational potential of each black hole. The second line in Eqs.~\eqref{dotc1} and ~\eqref{dotc2} represents the level mixing between modes of the same black hole, while the third line in each equation represents the mass transfer between different black holes.

\subsection{A two-mode model}
\label{sec:toymodel}

For simplicity, we shall assume the binary black holes moving in a circular orbit with their spins perpendicular to the orbital plan, and consider a two-mode model with each mode associated with different black holes. We choose the coordinates centered on the primary black hole with the $z$ axis perpendicular to the orbital plan and the $x$ axis pointing to the secondary black hole at $t=0$. In this case, Eqs.~\eqref{dotc1} and ~\eqref{dotc2} can be written as
\begin{equation}
i \begin{bmatrix}
    \dot{c}_{1} \\
    \dot{c}_{2} \\
\end{bmatrix}
=
\begin{bmatrix}
    A_{11} & A_{12}e^{i(\epsilon_2-\epsilon_1)t+i(m_2-m_1)\Phi} \\
    A_{21}e^{-i(\epsilon_2-\epsilon_1)t-i(m_2-m_1)\Phi} & A_{22} \\
\end{bmatrix}
\begin{bmatrix}
    c_1 \\
    c_2 \\
\end{bmatrix} ,
\label{matrix}
\end{equation}
where $\epsilon_a = \omega_a - \mu \simeq \mu /2n_a^2$ and $\Phi$ is the azimuth angle of the secondary black hole. $A_{ab}$ depends only on $R = |{\bf r}_1- {\bf r}_2|$, and is defined explicitly in Appendix~\ref{App:masstransfer}. The diagonal terms are the corrections to the eigenfrequencies $\omega_a$ due to the presence of another black hole. Thus, we can further define 
\be
\tilde{c}_1=c_1 e^{-i A_{11}t}, \quad  \tilde{c}_2=\sqrt{\frac{A_{12}}{A_{21}}}c_2 e^{-i A_{22}t},
\ee
with which, Eq.~\eqref{matrix} becomes
\be\ba
    i \dot{\tilde{c}}_1&=-\sqrt{A_{12}A_{21}}\, e^{i\epsilon_{21}t+i(m_2-m_1)\Phi}\, \tilde{c}_2, \\
    i \dot{\tilde{c}}_2&=-\sqrt{A_{12}A_{21}}\,e^{-i\epsilon_{21}t-i(m_2-m_1)\Phi}\, \tilde{c}_1 ,
    \label{eqeq}
\ea\ee
where $\epsilon_{21} \equiv \epsilon_2+A_{22} - \epsilon_1 - A_{11}$. As for initial conditions, we assume the cloud is initially hosted by the primary black hole, i.e., $c_1=1$ and $c_2=0$ as $t \rightarrow -\infty$. The orbital azimuthal angle can be written in the adiabatic limit,
\be
\Phi \simeq \pm \left( \Omega t + \dot{\Omega} t^2 \right),
\label{eq:Phi}
\ee
where the plus/minus sign corresponds to corotation/counterrotation case. If the orbit decays only by GW radiation, $\dot{\Omega}$ is given by Eq.~\eqref{eq:dOGR}. Given Eq.~\eqref{eqeq}, we expect that a resonance may happen and $c_1$ and $c_2$ when the orbital frequency near 
\be
  \Omega_{res}=\mp\frac{\epsilon_{21}}{m_2-m_1},
  \label{eq:reso}
\ee
which is the resonance frequency. Given the positivity of $\Omega$, we can have resonance only if $\mp \epsilon_{21}/(m_2-m_1)>0$. In addition, due to the symmetry of spherical harmonics $Y_{lm}(\theta,\phi)$, the couplings $A_{12}$ and $A_{21}$ are nonzero either with $m_{1}=l_{1}-2 \mathbf{Z}_1$ and $m_{2}=l_{2}-2\mathbf{Z}_2$, or with $m_{1} \neq l_{1}-2\mathbf{Z}_1$ and $m_{2} \neq l_{2}-2\mathbf{Z}_2$, where $\mathbf{Z}_1, \mathbf{Z}_2=0,1,2,\ldots$. For example, the cloud that is initially in the $|211\rangle$ mode can only transfer to the certain modes of the companion, such as $|21 -1\rangle$, $|31 -1\rangle$, $|320\rangle$,  $|322\rangle$, $|32-2\rangle$, etc.

Near the resonance, Eq.~\eqref{eqeq} can be approximated by
\begin{align}
    i \frac{{\rm d}\tilde{c}_1}{{\rm d}\tilde{t}}=-A\tilde{c}_2e^{\pm i\tilde{t}^{2}}, \quad   i \frac{{\rm d}\tilde{c}_2}{{\rm d}\tilde{t}}=-A\tilde{c}_1 e^{\mp i\tilde{t}^{2}},
    \label{simeq}
\end{align}
where we have defined $\tilde{t}=t\sqrt{(m_2-m_1) \dot{\Omega}}$ and $A=\sqrt{\frac{A_{21}A_{12}}{(m_2-m_1) \dot{\Omega}}}$. The general solution of Eq.~\eqref{simeq} can be written as
\begin{align}
    \tilde{c}_1&= {\rm C}_1  {\rm H} \left[-\frac{iA^2}{2},(-1)^{1/4} \tilde{t} \right] + {\rm C}_{2}  {\rm F} \left(\frac{iA^2}{4}, \frac{1}{2}, i\tilde{t}^2\right), \\
    \tilde{c}_2 &=-{\rm C}_1\ (-1)^{1/4}A e^{-i \tilde{t}^{2}} {\rm H} \left[-1-\frac{i A^{2}}{4},(-1)^{1/4} \tilde{t} \right]\nonumber\\
    &\,\,\,\,\, +i \, {\rm C}_2 \  A e^{-i \tilde{t}^2} \tilde{t}\ F \left(1+\frac{i A^2}{4};\frac{3}{2};i \tilde{t}^2\right), 
    \label{eq:ansolc2}
\end{align}
where $H(n,x)$ is the Hermite polynomials and $F(a,b,x)$ is the Kummer confluent hypergeometric function. Given the initial condition as $t \rightarrow - \infty$, we find 
\begin{equation}
    \begin{split}
    &{\rm C}_1=-e^{i\frac{3\pi}{4}-\frac{A^2 \pi}{8}},\  \\
    &{\rm C}_2=-\frac{1}{\pi}\Gamma\left(\frac{iA^2}{4}\right)\Gamma\left(1-\frac{iA^2}{2}\right)\sinh{\frac{A^2\pi}{2}}e^{i\frac{\pi}{4}-\frac{A^2\pi}{8}}.        
    \end{split}
\end{equation}
The cloud transfers to the secondary black hole after resonance can be found in the $t \rightarrow \infty$ limit,
\be
 |c_2^{\infty}|^2=  \frac{A_{21}}{A_{12}}\left(1-e^{-\pi A^2} \right) \simeq    \frac{\pi A_{21}^2}{{(m_2-m_1)\dot{\Omega}}} ,
 \label{prob}
\ee
where we considered $\pi A^2 \ll 1$. 

\begin{figure}
    \centering
    \includegraphics[height=0.87\linewidth]{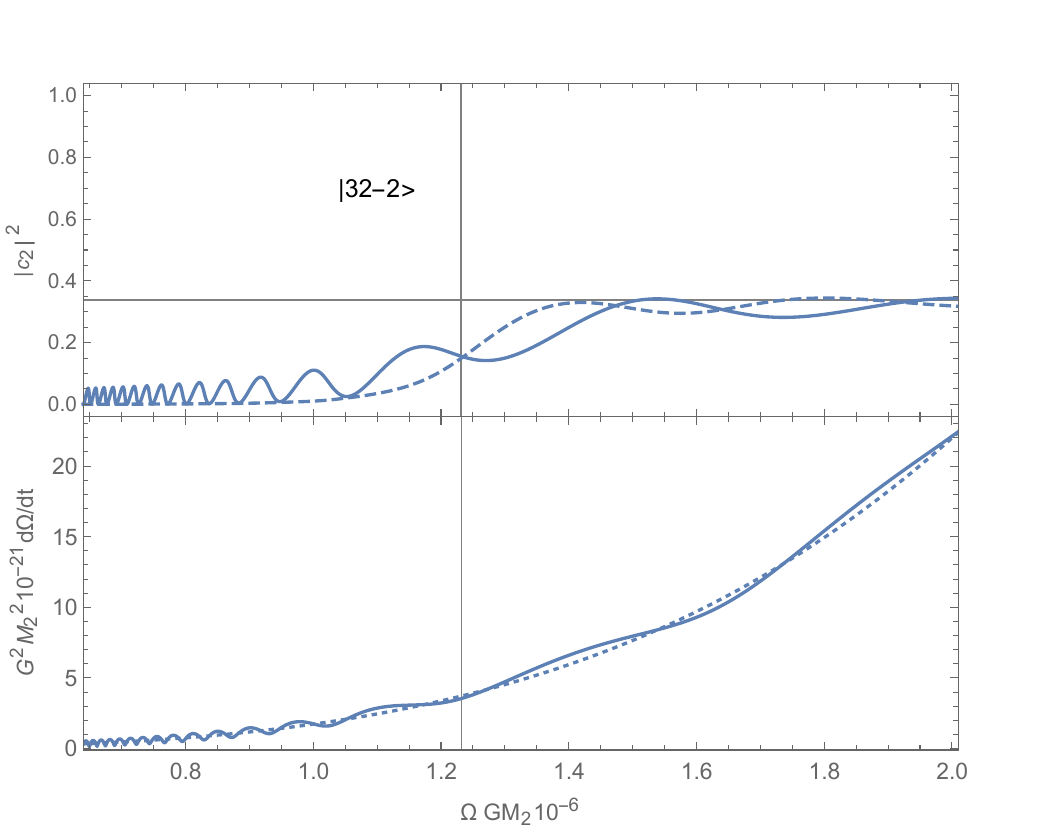}
    \caption{Mass transfer of cloud in a GW150914-like binary. We consider $\alpha_2 = 0.04$, and a cloud of $M_c = 0.1 M_1$ that is initially in the $|211\rangle$ mode of the primary black hole. The vertical gray line denotes the resonance frequency of transfer to the $|32-2\rangle$ mode of the secondary black hole. The upper panel shows the fractional mass transferred to the secondary black hole, where the solid blue line is obtained by numerically solving Eqs.~\eqref{eqeq} and~\eqref{dOmegaMT} with $\dot{\Phi} =\Omega(t)$, the dashed blue line shows the analytical solution~\eqref{eq:ansolc2}, and the gray dashed lines show the asymptotical value estimated by Eq.~\eqref{prob}. In the lower panel, the blue line shows $\dot{\Omega}$ given by the numerical solution as in the upper panel, while the dotted blue line is given by Eq.~\eqref{eq:dOGR}. The differences between two lines reflect the backreaction of mass transfer on orbital evolution.
}
    \label{fig:orbit}
\end{figure}

\subsection{Backreaction on orbital decay}

In the presence of mass transfer, the binding energy of the orbit evolves as
\be
\frac{{\rm d} E_{\rm orb}}{{\rm d} t} = \frac{G M_1M_2}{2 R} \left[ \frac{\dot{R}}{R}+ \frac{\gamma \left(q-1\right)}{q}\right],
\label{eq:dE}
\ee
where $\gamma \equiv -\dot{M_1}/M_1 $ denotes the efficiency of mass transfer and we assumed $\dot{M_1}=-\dot{M_2}$. During the resonance, about $M_c |c_2^{\infty}|^2$ amount of cloud mass is transferred in a timescale of $1/\sqrt{(m_2-m_1)\dot{\Omega}}$; thus the mass transfer efficiency can be estimated as
\be
\gamma \sim \frac{M_c}{M_1}  |c_2^{\infty}|^2 \sqrt{(m_2-m_1)\dot{\Omega}}\simeq \frac{M_c}{M_1} \frac{\pi A_{21}^2}{ \sqrt{(m_2-m_1)\dot{\Omega}}}\, .
\ee
By balancing Eq.~\eqref{eq:dE} with power of GW radiation, we find the orbital frequency evolution $\dot{\Omega}$ satisfying
\be
    \dot{\Omega}=  \frac{96}{5} \frac{\eta}{G^2 M_{\rm tot}^2} (GM_{\rm tot} \Omega)^{11/3} + \frac{3\gamma \left(q-1\right)}{2q}\Omega.
 \label{dOmegaMT}
\ee
For $q>1$, mass transfer of the cloud can accelerate the orbital decay. For $q<1$, the mass transfer can slow down the orbital decay. In particular, we write $\dot{\Omega} = \dot{\Omega}_{0} + \dot{\Omega}_{\gamma}$ with $\dot{\Omega}_{0}$ given by Eq.~\eqref{eq:dOGR} being the frequency evolution without mass transfer. For $\dot{\Omega}_{\gamma} \ll \dot{\Omega}_{0}$ we have $\dot{\Omega}_{\gamma}/\dot{\Omega}_{0} \simeq B$, while for $\dot{\Omega}_{\gamma} \gg \dot{\Omega}_{0}$ we have $\dot{\Omega}_{\gamma}/\dot{\Omega}_{0} \simeq B^{2/3}$, where
\be
B  \equiv \frac{5\sqrt{5}\pi  q^{-5/2} (q-1) }{256\sqrt{6(m_2-m_1)}} \frac{M_c}{M_1}  \frac{\left(GM_1 A_{21}\right)^2}{ \left(G M_1 \Omega \right)^{9/2}}.
\ee

In order to illustrate the backreaction of mass transfer, we evolve a GW150914-like system with its orbit sweeping over the resonant orbit of mass transfer. Instead of assuming Eq.~\eqref{eq:Phi}, we numerically solve Eqs.~\eqref{eqeq} and~\eqref{dOmegaMT} with $\dot{\Phi} =\Omega(t)$ for $c_{1,2}(t)$, $\Phi(t)$, and $\Omega(t)$. The results are shown in Fig.~\ref{fig:orbit}. As we expect, we find that the fractional mass transferred to the second black hole increases significantly when the orbital frequency matches the resonance frequency defined in Eq.~\eqref{eq:reso}, and eventually approaches to a value that agrees well with the analytical esitimation, Eq.~\eqref{prob}. We also find the backreaction on the orbital evolution, at least in the considered case, is small (cf. the lower panel in Fig.~\ref{fig:orbit}). 
Moreover, we find mass transfer also introduces sloshing oscillation in $\dot{\Omega}$ with a frequency given by $\sim \sqrt{A_{21}A_{12}}$ near the resonance.

\begin{figure}
    \centering
    \includegraphics[width=0.85\linewidth]{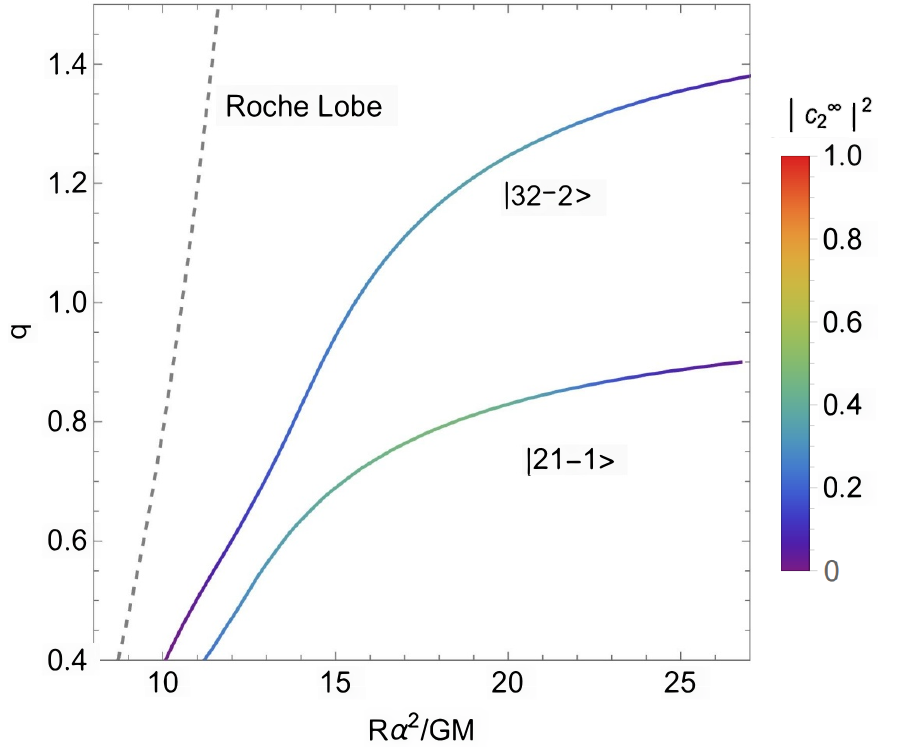}
    \caption{Resonance orbits of mass transfer from the $|211\rangle$ mode of one black hole to the $|21-1\rangle$ and $|31-2\rangle$ modes of another black hole. Color on the curve reflects the square of the probability amplitude of the transferred cloud, assuming $M_c = 0.1 M$ before the mass transfer. The gray dashed line shows the Roche lobe~\cite{Eggleton:1983rx}. Notice that $q=M_*/M$ here is defined as the mass ratio of the cloud receiver divided by the cloud donor, and the result
shown in the plot does not depend on the cloud mass. }
    \label{fig:mt}
\end{figure}

Assuming the cloud is initially in the $|211\rangle$ mode of its host black hole in a corotating binary, we investigate its transfer to the $|21-1\rangle$ and $|32-2\rangle$ modes of the companion black hole, respectively, by calculating Eqs.~\eqref{eq:reso},~\eqref{dOmegaMT} and~\eqref{prob} numerically. The results are shown in Fig.~\ref{fig:mt}. We find that the transfer efficiency, captured by $|c_2^{\infty}|^2$, is suppressed by large $\dot{\Omega}$ if the resonance happens at small separation, and is suppressed by small $A_{21}^2$ if the resonance happens at large orbit.

Moreover, we find that the resonant transfer happens at larger radii than the Roche Lobe transfer, i.e., at earlier evolution time. In addition, for a given type of binary, the transferred cloud mass is a function of the ALP wavelength, which maximizes around certain values. Third, for a binary with a given mass ratio, it is possible that we need to consider resonant mode transfer to multiple target states.  Interestingly, the cloud receiver can be a neutron star as the $|211\rangle \rightarrow |21-1\rangle$ channel is viable for small mass ratios. A cloud can transfer from a black hole to a neutron star and induce transient electromagnetic radiations if the cloud consists of particles coupling with the electromagnetic sector in the standard model. This is an interesting process that is worth further studies.

\begin{figure}
    \centering
    \includegraphics[height=0.8\linewidth]{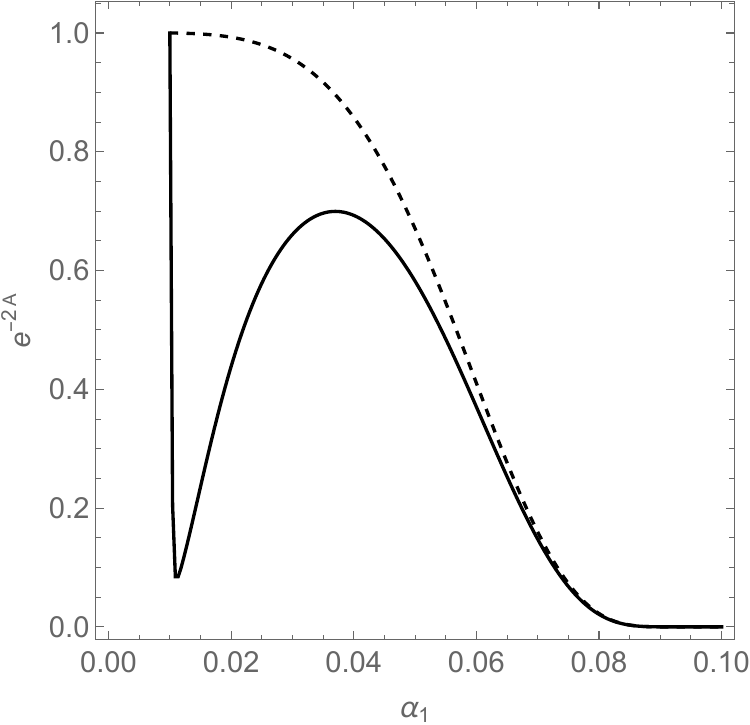}\\
     \includegraphics[height=0.8\linewidth]{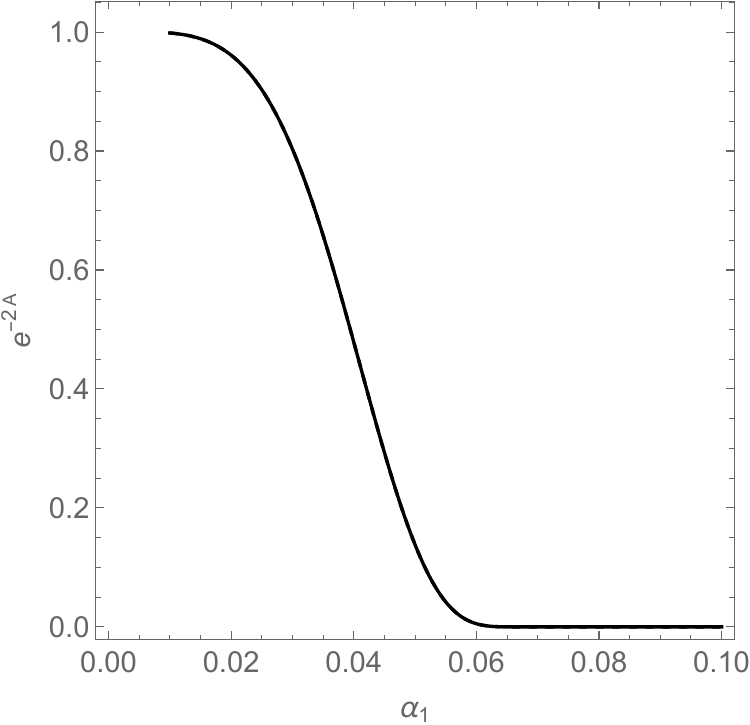}
    \caption{Parameter space for cloud around the primary black hole (upper panel) and secondary black hole (lower panel) in a GW151226-like binary to survive to mass transfer to the $|21-1\rangle$ mode of their companions. We consider level mixing with the $|21-1\rangle$, $|311\rangle$, $|100\rangle$, and $|200\rangle$ modes. The solid lines show the fractional cloud mass at the mass transfer. We also show the results obtained by considering level mixing only to the $|21-1\rangle$ mode, in dashed lines. In addition, we find that, for $\alpha_1 < 0.009$, the binary will sweep the mass transfer orbit during the CE evolution.}
    \label{fig:dpmt}
\end{figure}

\begin{figure}
    \centering
    \includegraphics[height=0.8\linewidth]{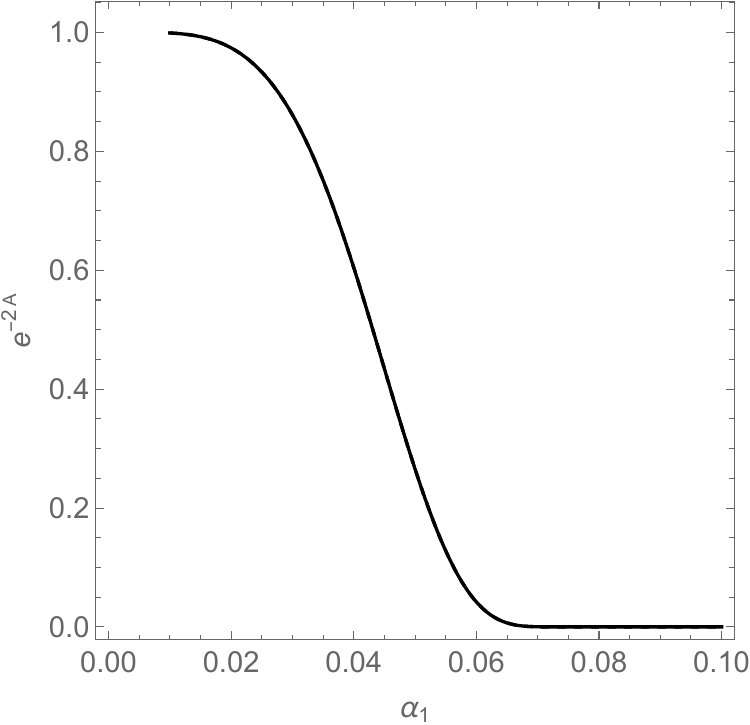}\\
     \includegraphics[height=0.8\linewidth]{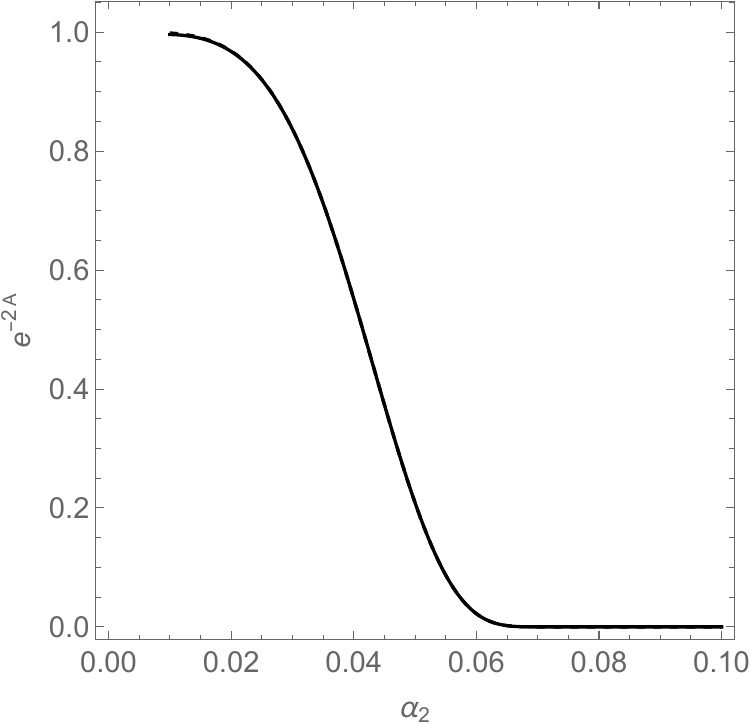}
    \caption{Parameter space for cloud around the primary black hole (upper panel) and secondary black hole (lower panel) in a GW150914-like binary to survive to mass transfer to the $|21-1\rangle$ mode of their companions. We consider level mixing with the $|21-1\rangle$, $|311\rangle$, $|100\rangle$, and $|200\rangle$  modes. The solid lines show the fractional cloud mass at the mass transfer. We also show the results obtained by considering level mixing only to the $|21-1\rangle$ mode, in dashed lines. In addition, we find that, for $\alpha_1 < 0.007$, the binary will sweep the mass transfer orbit during the CE evolution.}
    \label{fig:dpmt0914}
\end{figure}

In the end of this section, we would like to investigate the parameter space for superradiant cloud mass transfer. In particular, we would like to know, in typical binary evolution, how much cloud can survive from level mixing depletion so that they can undergo mass transfer. Again, we consider the GW151226-like and the GW150914-like binaries, and perform the integral~\eqref{eq:dep1} from the formation of the host black hole to the time when the binary reaches the mass transfer resonance orbit. The results are shown in Figs.~\ref{fig:dpmt} and~\ref{fig:dpmt0914}. We know that, when a GW151226-like binary reaches the resonance orbit of mass transfer, the $|211\rangle$ mode that grew around the primary black hole almost depleted due to its mixing with the $|21-1\rangle$ mode if $\alpha_1 > 0.07$ and almost depleted due its mixing to the $|200\rangle$ mode if $\alpha_1 < 0.02$. For a cloud that grew around the secondary black hole, it almost depleted due to its mixing to the $|21-1\rangle$ if $\alpha_2 > 0.05$ when the binary reaches the resonance orbit, otherwise the cloud can survive and transfer to the other black hole in the binary. For a GW150914-like binary, cloud that grew around the primary black hole could avoid level mixing depletion and survive to mass transfer if $\alpha_1 < 0.05$. The cloud that grew around the secondary black hole can avoid level mixing and transfer to another black hole if $\alpha_2<0.06$.

\section{Discussion and Conclusion}

This study provides a natural explanation for the coexistence of superradiant clouds and close compact binaries, and highlights the importance of fully incorporating relevant astrophysical processes when we design and perform the tests for ultralight bosons beyond the standard model, as the actual astrophysical environmental effects are often nonperturbative with respect to purely gravitational considerations. Here the key insight comes from the fact that in the astrophysical evolution path of close compact binaries, the gas-binary interaction during the CE phase brings the binaries to much closer distances within thousands of years, which is much shorter than the gravitational radiation reaction timescale. As a result, some level mixings may no longer act as an effective damper for the clouds as previously considered. By including all the main damping mechanisms of the clouds into the evolution model of the compact binaries, for sample binary systems we have computed the level of cloud growth and depletion for various orbital frequencies and particle masses. It is evident that the CE process does enlarge the parameter space that clouds coexist with the compact binary within the detection band of space-borne gravitational wave detectors such as LISA. This is interesting because the compact binaries carrying superradiant clouds may evolve differently from those without, because of the quadrupole moment of the clouds. This also means that the gravitational waveform will be significantly modified and potentially detectable by space-borne gravitational wave detectors, if some of them were within the horizon of these detectors. For sources within the band of ground-based detectors, however, the level mixing may severely deplete the clouds before they evolve to such a close distance due to gravitational wave radiation. Additional studies are needed to analyze the possibility of cloud transfer to circum-binary states, as compared to the circum-single states considered here. The circum-binary states are analogous to the molecule states, within which the ultralight particles orbit around the binary instead of mainly around the individual compact object. If the transfer to the circum-binary state is effective before the total depletion of the clouds due to level mixing, this may also lead to the possibility of close compact binaries in the LIGO-Virgo-KAGRA band while the inteaction with clouds is still dynamically important. This scenario is certainly worth more exploration given the rapid growing catalog of compact binaries observed by Advanced LIGO and Virgo. 

During the evolution of these close compact binaries we also find that clouds may resonantly transfer between compact binary objects. This happens because the circum-single cloud states around two different objects may have a small overlap with each other. If the binary orbital frequency matches the frequency difference between these two states, enabling resonant transfer that build up in time.
This mass transfer mechanism relies crucially on the wave nature of the clouds, so that it is clearly different from the Roche-Lobe filling mass transfer, which generally occurs at different orbital separations. Interestingly, when the receiver is a neutron star, the received particles such as ALPs may interact with the strong magnetic field of the neutron star if their coupling strength is nonzero, and produce electromagnetic counterparts, even if the neutron star cannot superradiantly excite the axion cloud efficiently.\footnote{Superradiant generation of ALPs by neutron stars is possible with additional coupling to the electromagnetic sector \cite{Day:2019bbh}.}A Similar effect in terms of dark photon interacting with electromagnetic fields around black holes has been considered in Ref.\cite{siemonsen2023dark}. The detailed observational signatures depend on the nature of the ALPs,  the coupling to the Standard Model sector, and the environment. Understanding the detailed electromagnetic signatures and developing associated searching strategies will be another interesting direction to explore. While focusing on clouds of scalar fields in this work, we expect that the common envelop process as well as the cloud mass transfer process are also relevant for clouds of vector fields. The details require further investigation and this is left for future study.

For compact binaries detectable by ground-based detectors, another viable formation channel is through dynamical few-body interaction in dense stellar environments, e.g., globular clusters and nuclear star clusters \cite{Rodriguez:2015oxa,Fragione:2018vty,Antonini:2016gqe,Banerjee:2016ths}. Recently, there was also extensive exploration of stellar-mass compact object evolution and binary formation in active Galactic nucleus, which are viable sources for both ground-based and space-borne gravitational-wave detection \cite{Sigl:2006cg,McKernan:2012rf,Bartos:2016dgn,Stone:2016wzz,Tagawa:2019osr,McKernan:2020lgr,Levin:2003ej,Levin:2006uc,Pan:2021ksp,Pan:2021oob,Pan:2021xhv} and possible origin of environmental signals in the gravitational waveform \cite{Yunes:2011ws,Barausse:2014tra,Bonga:2019ycj,Sberna:2022qbn,LISAConsortiumWaveformWorkingGroup:2023arg}.
For these dynamically formed binaries, the vastly different evolution path likely lead to distinct cloud evolution histories as well, as the depletion due to hyperfine resonance may be alleviated by the fast dynamical evolution of orbital separation. It is important to  track the detailed evolution in each dynamical channel, accounting for the relevant cloud damping mechanisms and address whether the dynamical channels also lead to significant cloud presence in the gravitational wave bands. 

\section*{ACKNOWLEDGMENTS}

We thank Junwu Huang for helpful discussions. J. Z. is supported by the scientific research starting grants from University of Chinese Academy of Sciences (Grant No.~118900M061), the Fundamental Research Funds for the Central Universities (Grants No.~E2EG6602X2 and No.~E2ET0209X2), and the National Natural Science Foundation of China (NSFC) under Grant No.~12147103. H. Y. is supported by the Natural Sciences and Engineering Research Council of Canada and in part by Perimeter Institute for Theoretical Physics. Research at Perimeter Institute is supported in part by the Government of Canada through the Department of Innovation, Science and Economic Development Canada and by the Province of Ontario through the Ministry of Colleges and Universities. 

\newpage

\appendix

\section{LEVEL MIXING IN GRAVITATIONAL ATOM}
\label{App:levelmixing}

In this Appendix, we review the level mixing of a cloud induced by the gravitational potential of the companion object. While previous studies usually assume the companion object is far away from the cloud, i.e., $R \gg r_g/\alpha^2$, we extend the calculation to the case that the companion immerses in the cloud. In particular, we emphasize the mixing via the $\ell = 0, 1$ multipoles, as well as nonresonant cloud depletion.

In the frame centered at the superradiance black hole, the Hamiltonian of the system is
\be\ba
H &= \left[\frac{{\bf p}^2}{2\mu} - \frac{G\mu M}{r} \right]  \\
&+ \left[ \frac{{\bf p}_*^2}{2M_*} - \frac{GM_*M}{R} - \frac{GM_*\mu}{\left| {{\bf r}_*} - {\bf r} \right|} + \frac{GM_*\mu}{R^3} {\bf r}\cdot {{\bf r}_*}\right], 
\label{eq:Htot}
 \ea\ee
where ${\bf p} = \mu \dot{{\bf r}}$ and ${\bf p}_* = M_* \dot{\bf r}_*$ with ${\bf r}$ and ${\bf r}_*$ being the positions of the cloud and the star relative to the black hole~\cite{Baumann:2018vus}. Therefore, the companion provides an additional gravitational potential,
\be
V_*=- \frac{GM_*\mu}{\left| {{\bf r}_*} - {\bf r} \right|} + \frac{GM_*\mu}{R^3} {\bf r}\cdot {{\bf r}_*},
\ee
where the second term appears become the frame centred at the superradiance black hole is not an inertial frame. For simplicity, we shall consider the two-mode toy model, in which the wave function of the cloud can be written as 
\be
\psi =c_d(t)\, \psi_d +c_g(t)\, \psi_g
\ee
with $\psi_g$ and $\psi_{d}$ being the growing and decaying modes respectively. Specifically, $\psi_g$ and $\psi_{d}$ take the form of
\begin{equation}
    \psi_{n\ell m} =R_{n\ell}(r)Y_{\ell m}(\theta,\phi)e^{-i(\omega_{n\ell m}-\mu)t},
    \label{wavefunction}
\end{equation}
with
\begin{equation}
\begin{aligned}
\omega_{n\ell m}\simeq \mu &\left[1-\frac{\alpha^2}{2n^2}-\frac{\alpha^4}{8n^4} +\frac{\left(2\ell-3n+1\right)\alpha^4}{n^4\left(\ell+1/2\right)}\right. \\
&\left.+\frac{2a m\alpha^5}{n^3\ell (\ell+1/2)(\ell+1)}\right].  
\end{aligned}
\end{equation}
The coefficients satisfy
\begin{equation}
i \frac{d}{dt}\begin{bmatrix}
    c_{g}(t) \\
    c_{d}(t) \\
\end{bmatrix}
=
\begin{bmatrix}
    0 & \langle\psi_g|V_*|\psi_d\rangle \\
    \langle\psi_d|V_*|\psi_g\rangle& 0 \\
\end{bmatrix}
\begin{bmatrix}
    c_g(t) \\
    c_d(t) \\
\end{bmatrix},\label{eq:cgcd}
\end{equation}
where we have absorbed the diagonal terms $\langle\psi_g|V_*|\psi_g\rangle$ and $\langle\psi_d|V_*|\psi_d\rangle$ in $\omega_g$ and $\omega_d$. Given the off-diagonal terms, we can define the resonant orbital frequency,
\be
\Omega_0 \equiv \frac{\omega_g -\omega_d}{m_g-m_d},
\ee
and the coupling strength $\eta$ by
\be
\langle\psi_g|V_*|\psi_d\rangle \equiv \eta e^{-i (m_d-m_g) \left(\Omega_0 \mp \Omega\right) t},
\ee
where the minus and plus signs correspond to corotating and counterrotating orbits, respectively.

The coupling strength can be calculated by performing the multipole expansion of $V_*$. For $r < R$, we have
\be\ba\label{eq:1overR}
 \frac{1}{\left| {{\bf r}_*} - {\bf r} \right|}  &= \frac{1}{R} + \frac{r \cos \Delta\theta}{R^2}   \\ 
&+  \sum_{\ell \ge 2}\, \sum_{\left|m \right| \le \ell} \frac{4\pi}{2\ell+1} \frac{r^{\ell}}{R^{\ell+1}} Y^*_{\ell m}\left(\theta_*,\phi_*\right) Y_{\ell  m}\left(\theta,\phi\right),
\ea\ee
where $\Delta \theta$ is the angle between ${\bf r}$ and ${\bf r}_*$. The first term on the rhs of Eq.~\eqref{eq:1overR} does change the eigenstate of the cloud as it is a constant for $r < R$. Its contribution would be clear when we calculate $\I_r$ defined later. The second term cancels with the last term in Eq.~\eqref{eq:Htot}. For $r > R$, we have
\be\ba
 \frac{1}{\left| {{\bf r}_*} - {\bf r} \right|}  &= \frac{1}{r} + \frac{{\bf r}\cdot {{\bf r}_*}}{r^3}   \\
 &+\sum_{\ell \ge 2} \,\sum_{\left|m\right| \le \ell} \frac{4\pi}{2\ell+1}  \frac{R^{\ell}}{r^{\ell+1}} Y^*_{\ell m}\left(\theta_*,\phi_*\right) Y_{\ell  m}\left(\theta,\phi\right),
\ea\ee
in which case the monopole and dipole terms contribute. With the multipole expansion, The inner product can be written as
\be\ba
\langle\psi_i|V_*|\psi_j\rangle &=& -GM_* \mu \sum_{\ell,\, m} \frac{4\pi}{2\ell + 1} \,\I_\Omega\, \I_r(R) Y_{\ell m}(\theta_*,0) \nonumber \\
&&\times \exp \left[ i(\omega_i-\omega_j)t\mp im \phi_*\right],
\ea\ee
where $i, j\in\{g, d\}$ and
\begin{equation}
        \I_{\Omega}=\int d\Omega Y^{*}_{\ell_{j} m_{j}}(\theta,\phi)Y_{\ell_{i}m_{i}}(\theta,\phi)Y_{\ell m}(\theta,\phi).
\end{equation}
The upper/lower sign in front of $m\phi_*$ corresponds to corotating/counterrotating orbits. For $\ell =0$ and $\ell \ge 2$, we have
\be\ba
\I_{r}=\int_0^{\infty} dr \, r^2 \frac{r_{<}^{\ell}}{r_{>}^{\ell+1}} R_{n_i \ell_i}\left(r\right) R_{n_j \ell_j}\left(r\right),
\ea\ee
where $r_{<}$ is the smaller of $r$ and $R$ and $r_{>}$ is the larger of $r$ and $R$. For $\ell_* = 1$, we have
\be\ba
\I_{r} = \int_{R}^{\infty} dr \, R \left(1- \frac{r^3}{R^3}\right) R_{n_i \ell_i}\left(r\right) R_{n_j \ell_j}\left(r\right).
\ea\ee
Therefore, we find that terms with $\ell=0,1$ will be exponentially suppressed by the radial function if $R \gg r/\alpha^2$, but could contribute significantly if $R \lesssim r/\alpha^2$. Moreover, the integral $\I_{\Omega}$ implies the following selection rules: 
\begin{equation}
    \left\{
\begin{aligned}
   &-m+m_i+m_j=0 ,\\
   & |\ell_i-\ell_j|\le \ell \le \ell_i+\ell_j, \\
   & \ell+\ell_i+\ell_j=2k, \quad {\rm for} \  k\in\mathbb{Z}. \\
\end{aligned}
\right.
\end{equation}
Based on the rules above, inner products between the growing mode $|211\rangle$ and decaying modes with $\ell =m = 0$ give nonzero, and can lead to depletion channels in addition to the hyperfine and Bohr mixing studied in Ref.~\cite{Baumann:2018vus}.

Now we discuss cloud depletion induced by level mixing. Near the resonant orbit, i.e., $\Omega \sim \Omega_0$, the orbit decay should be described by $\Omega = \Omega_0 + \gamma t$, and we can find a resonant solution known as the Landau-Zener transition~\cite{baumann2020gravitational}. While away from the resonant orbit, e.g., $\Omega \ll \Omega_0$, we can take the adiabatic approximation and consider $\Omega$ to be a constant, then we have
\begin{equation}
\begin{aligned}
    |c_{d}(t)|^2&=\sin^2\left[\int^t_{t_0}dt'\sqrt{\eta^2+\left(\Omega_0 \mp \Omega(t')\right)^2}\right]\\
    &\times\left[1-\frac{\left(\Omega_0 \mp \Omega(t)\right)^2 }{\eta^2+\left(\Omega_0 \mp \Omega(t)\right)^2}.
    \right],
    \label{eq:c2dep}
    \end{aligned}
\end{equation}
In this case, cloud depletion can be estimated with Eq.~\eqref{eq:dep1}. We replace the oscillating term in Eq.~\eqref{eq:c2dep} with a factor of $1/2$ for simplicity when performing the numerical integration. We find that the nonresonant solution also leads to a significant depletion of the cloud.

Level mixing also modifies $\omega_{g}$ and $\omega_{d}$. Under the WKB approximation, the modifications are~\cite{Tong:2022bbl},
\begin{equation}
 \begin{aligned}
   &\delta \omega_{g} =\frac{|\langle\psi_g|V_*|\psi_d\rangle|^2}{\omega_g-\omega_d}+i\left[\Gamma_g-\frac{\Gamma_g-\Gamma_d}{(\omega_g-\omega_d)^2}|\langle\psi_g|V_*|\psi_d\rangle|^2\right], \nn \\
   &\delta \omega_{d} =\frac{|\langle\psi_g|V_*|\psi_d\rangle|^2}{\omega_d-\omega_g}+i\left[\Gamma_d-\frac{\Gamma_d-\Gamma_g}{(\omega_g-\omega_d)^2}|\langle\psi_g|V_*|\psi_d\rangle|^2\right]. 
\end{aligned}
\end{equation}
As a result, the superradiance rate of a growing mode coupled to a decaying mode acquires a suppression
\begin{equation}
    \Gamma_{g}'=\Gamma_g+\Delta \Gamma_g\simeq\Gamma_g-\frac{\Gamma_g-\Gamma_d}{(\omega_g-\omega_d)^2}|\langle\psi_g|V_*|\psi_d\rangle|^2<\Gamma_g,
\end{equation}
where $\Gamma_g$ is the original growth rate defined in Eq.(\ref{growrate}). In particular, the superradiance rate $\Gamma_{d}$ could become negative even at maximal black hole spin when the orbital separation reaches to some critical value $R_{\rm cr}$. By considering the case of $r > R$ when calculating $\langle\psi_{g}|V_*|\psi_{d}\rangle$, we find that modes with $\ell=0$ also contribute to $R_{\rm cr}$, and extend the results in Ref.~\cite{Tong:2022bbl} to the region with $R_{cr} \lesssim r_g/\alpha^2$.

\section{COUPLING STRENGTH BETWEEN BOUND STATES ASSOCIATED WITH DIFFERENT BLACK HOLES}
\label{App:masstransfer}

In the two-modes toy investigated in Sec.~\ref{sec:toymodel}, Eqs.~\eqref{dotc1} and~\eqref{dotc2} reduce to Eq.~\eqref{matrix} with matrix elements defined by

\begin{align}
&A_0=1-\langle\psi^*_{2}\psi_{1}\rangle\langle\psi^*_{1}\psi_{2}\rangle,\nn\\
&A_{11}=\frac{ \langle\psi^*_{1}V_{2}\psi_{1}\rangle-\langle\psi^*_{1}\psi_{2}\rangle\langle\psi^*_{2}V_{2}\psi_{1}\rangle}{A_0},\nn\\
 &A_{22}=\frac{  \langle\psi^*_{2}V_1\psi_{2}\rangle-\langle\psi^*_{2}\psi_{1}\rangle\langle\psi^*_{1}V_1 \psi_{2}\rangle }{A_0},\nn\\
  &A_{12}=  \frac{ \langle\psi^*_{1}V_1\psi_{2}\rangle-\langle\psi^*_{1}\psi_{2}\rangle\langle\psi^*_{2}V_1 \psi_{2}\rangle}{A_0}e^{-i(\epsilon_2-\epsilon_1)t-i(m_2-m_1)\phi^*},\nn\\
 &A_{21} = \frac{\langle\psi^*_{2}V_{2}\psi_{1}\rangle-\langle\psi^*_{2}\psi_{1}\rangle\langle\psi^*_{1}V_{2}\psi_{1}\rangle}{A_0}e^{-i(-\epsilon_2-\epsilon_1)t+i(m_2-m_1)\phi^*}.\nn
\end{align}

In this Appendix, we calculate the coupling strength between bound states associated with different black holes. We shall work in a coordinate frame $\{\bar{r}, \bar{\theta}, \bar{\phi} \}$, which is centered at the primary black hole with the $z$ axis aligning with the spin of the primary black hole and the $x$ axis always pointing to the secondary black hole. Considering the coordinate frames centered at the two black holes, respectively, say $\{r_1, \theta_1, \phi_1 \}$ and $\{r_2, \theta_2, \phi_2 \}$, with their $z$ axes aligning with the black hole spin (here we only consider the case with parallel black hole spins), then we have  
\begin{equation}
    \{ r_1, \theta_1, \phi_1 \} = \{\bar{r}, \bar{\theta}, \bar{\phi}+\phi^* \},
\end{equation}
and
\begin{equation}
    \begin{aligned}
    \{& r_2, \theta_2, \phi_2 \} \\
    = &\left\{  \sqrt{\bar{r}^2\sin^2{\bar{\theta}}\sin^2{\bar{\phi}}+r^2\cos^2{\bar{\theta}}+(R-r\sin{\bar{\theta}}\cos{\bar{\phi}})^2}, \right.\\
    & \tan^{-1}\left[\sqrt{R^2 - 2 \bar{r} R \cos \bar{\phi} \sin \bar{\theta} + \bar{r}^2 \sin^2{\bar{\theta}}}\Big/\bar{r}\cos\bar{\theta}\right],\\
    & \left. \tan^{-1}\left[\bar{r}\sin{\bar{\theta}}\sin{\bar{\phi}}\big/ \left(\bar{r}\sin{\bar{\theta}}\cos{\bar{\phi}}-R\right)\right]+\phi^* \right\}.
\end{aligned}
\end{equation}
where $\phi^*=\pm\int \Omega(t)dt \simeq \pm \Omega t$ with $\pm$ corresponding to the corotating and counterrotating system respectively. Then we can calculate the coupling strength in $\{\bar{r}, \bar{\theta}, \bar{\phi} \}$. For example, 
\begin{equation}
\begin{split}
     \langle\psi^*_{1} \psi_{2} \rangle&=e^{i(\epsilon_2-\epsilon_1)t}\int d\bar{r} d\bar{\theta} d\bar{\phi} \  \bar{r}^2\sin{\bar{\theta}}\\
     &\times R^*_{n\ell}(r_1)Y^*_{\ell m}(\theta_1,\phi_1)R_{n'\ell'}(r_2)Y_{\ell'm'}(\theta_2,\phi_2),
\end{split}
\end{equation}
which can be calculated numerically.

\bibliography{references}

\end{document}